\documentclass[useAMS,usenatbib]{mn2e}

\usepackage{epsfig}
\usepackage{longtable}
\usepackage{times}
\usepackage{deluxetable}

\def \inte {$INTEGRAL$}
\def \xmm {$XMM-Newton$}

\def \sw {$Swift$}
\def \suz {$Suzaku$}
\def \src {IGR\,J08408--4503}

\def \hcm {\hbox {\ifmmode $ atom cm$^{-2}\else atom cm$^{-2}$\fi}}

\def \arcsec {\hbox{$^{\prime\prime}$}}
\def \chisq {$\chi ^{2}$}

\def \ATel {Astron.\ Tel.}
\def \apj {ApJ}
\def \apjl {ApJL}
\def \apjs {ApJS}
\def \aap {A\&A}

\def \pasj {PASJ}

\def \mnras {MNRAS}

\title[\suz\ observes \src\ during a long low luminosity state]{The longest observation of a low intensity state from a Supergiant Fast X--ray Transient: \suz\ observes \src}
\author[L. Sidoli, et al.]{L.\ Sidoli,$^{1}$\thanks{E-mail: sidoli@iasf-milano.inaf.it} P.\ Esposito,$^{2,3}$ L.\ Ducci,$^{4,1}$ \\
$^{1}$INAF, Istituto di Astrofisica Spaziale e Fisica Cosmica,
	Via E.\ Bassini 15,   I-20133 Milano,  Italy \\
$^{2}$ INAF, Osservatorio Astronomico di Cagliari, localit\`a Poggio dei Pini, 
strada 54, I-09012 Capoterra, Italy\\
$^{3}$ INFN, Istituto Nazionale di Fisica Nucleare, sezione di Pavia, via 
A.~Bassi 6, I-27100 Pavia, Italy\\
$^{4}$Dipartimento di Fisica e Matematica, Universit\`a dell'Insubria, Via
Valleggio 11, I-22100 Como, Italy \\
}

\begin{document}

\date{Accepted 2010 July 6.  Received 2010 July 6; in original form 2010 June 9}

\pagerange{\pageref{firstpage}--\pageref{lastpage}} \pubyear{2010}

\maketitle

\label{firstpage}

\begin{abstract}

We report here on the longest deep X--ray observation of a  supergiant fast X--ray transient (SFXT) outside outburst, 
with an average luminosity
level of  10$^{33}$~erg~s$^{-1}$ (assuming 3~kpc distance).
This observation was performed with \suz\  in December 2009 and was targeted on \src, 
with a net exposure with the  X--ray imaging spectrometer (XIS, 0.4--10 keV) 
and the hard X--ray detector  (HXD, 15--100 keV) 
of 67.4 ks and 64.7 ks, respectively, 
spanning about three days. 
The source was caught in a low intensity state characterized by an initially 
 average X--ray luminosity level  of 4$\times$10$^{32}$~erg~s$^{-1}$ (0.5-10 keV) during 
the first 120~ks,
followed by two long flares (about 45~ks each) peaking at a flux a factor
of about 3 higher than the initial pre-flare emission.
Both XIS spectra (initial emission and the two subsequent long flares) can be fitted with
a double component spectrum, with a soft thermal plasma model together with a power law, differently absorbed.
The spectral characteristics suggest that the source is accreting matter even at this very low intensity level.
From the HXD observation we place an upper limit 
of 6$\times$10$^{33}$~erg~s$^{-1}$ (15--40 keV; 3 kpc distance)
to the hard X--ray emission, which is the most stringent constraint on the hard X--ray emission 
during a low intensity state in a SFXT, to date.
The timescale observed for the two low intensity long flares 
is indicative of an orbital separation of the order of 10$^{13}$~cm in \src. 

\end{abstract}


	\section{Introduction\label{igr08408:intro}}

\src\ is a hard X--ray transient discovered in 2006 May with \inte\ 
during a  flare with a duration of 900~s, with  a flux of 250 mCrab (at peak, 20--40 keV) 
\citep{Gotz2006:08408-4503discovery}. 
Archival \inte\ observations of this field before the discovery observation 
demonstrated the recurrence of its 
flaring activity \citep{Mereghetti:08408-4503}.
A refined X--ray position \citep{Kennea2006:08408-4503},
allowed the association of 
the X--ray transient  with a bright O8.5Ib(f) star (HD~74194) located 
in the Vela region,  at a distance of about 3\,kpc \citep{Masetti2006:08408-4503}.
Optical spectroscopy of HD~74194, performed just a few days after the source discovery, 
showed variability in the H${\alpha}$ profile and a radial velocity variation 
with an amplitude of about  35 km\,s$^{-1}$ (He\,I and He\,II absorption lines; \citealt{Barba:2006}).
Other bright X--ray flares were observed with \inte\ and \sw\ 
in October 2006, July 2008 and September 2008
(\citealt{Gotz2007:08408-4503},  
\citealt{Romano2009:sfxts_paper08408}, 
\citealt{Sidoli2009}).
An orbital period of $\sim$35 days has been suggested based on the duration 
of the flaring activity \citep{Romano2009:sfxts_paper08408},
although a secure determination  based on X--ray and/or optical modulation is still lacking. 
Also a pulse period has never been found, thus the nature of the compact object is unknown.

\src\ has been classified as a member of the class of 
high mass X--ray binaries (HMXBs) called supergiant fast X--ray transients (SFXTs), 
X--ray transients with bright and short duration flaring activity
associated with blue supergiant companions 
(\citealt{Sguera2005}, \citealt{Smith2006aa}). 
The large majority of the members of this class have been discovered 
by \inte\ during the survey of the Galactic plane 
(\citealt{Sguera2005}, \citealt{Sguera2006}).
A few SFXTs are X--ray pulsars, thus demonstrating that 
the compact object is a neutron star,
while in the other sources a black hole cannot be excluded, although the
broad band spectra (0.1-100 keV) in outburst are very similar to
those typically observed in accreting pulsars (e.g. \citealt{Romano2008:sfxts_paperII}).
Their outbursts typically show a duration of a few days 
(\citealt{Romano2007}, \citealt{Sidoli2008:sfxts_paperIII}), with small duty cycles
although highly variable from source to source \citep{Romano2009:sfxts_paperV}.
Each outburst is characterized by several short (a few minutes to a few hour duration) 
and bright flares, each reaching a luminosity of a few 10$^{36}$--10$^{37}$~erg~s$^{-1}$
\citep{Sidoli2008:sfxts_paperIII}.
A much lower intensity emission characterizes the long term behaviour of these transients
(10$^{33}$--10$^{34}$~erg~s$^{-1}$, \citealt{Sidoli2008:sfxts_paperI}), while the quiescent
state (no accretion) has been observed rarely and displays a very soft
(likely thermal) spectrum  with a luminosity of $\sim$10$^{32}$~erg~s$^{-1}$ 
(\citealt{zand2005}, \citealt{Leyder2007}).
A very low level of accretion ($\sim$10$^{32}$~erg~s$^{-1}$)  has been observed from \src\ 
in 2007 with \xmm\ as well,
displaying a double component spectrum composed of a thermal soft part
(likely from the supergiant companion), together with a harder power-law component (\citealt{Bozzo2010}).

Although SFXTs only sporadically undergo 
a bright outburst (from null to 4 outbursts per source have been observed during the
two years long  monitoring of four SFXTs with \sw, ~\citealt{Romano2009:sfxts_paperV}), 
their properties during outbursts are
much better studied than their behaviour during the long-term lower intensity states 
(intermediate level or the quiescence). 

We report here on a \suz\ observation  of \src\ with a net exposure of 67.4~ks (XIS), 
spanning about three days. 
This is the longest X--ray observation of a SFXT catching the source in 
a very low intensity level.

We also report on the analysis of
our private \inte\ observations of the source field covering about one month including times of the \suz\ observation.
We then compare our \suz\ results with a much shorter observation of a low intensity state from \src\
performed with \xmm\ in 2007. 
To this aim, we  re-analysed these \xmm\ observations, although already reported by \citet{Bozzo2010}.

 	 \section{Observations and Data Reduction\label{igr08408:dataredu}}

\subsection{\suz\label{igr08408:suzredu}}

A \suz\ \citep{Mitsuda2007} observation of \src\ was performed 
from 2009 December 11, starting at 17:53 UT, until December 14 12:00 UT. 
The target was placed at the `HXD nominal' position. 
The X--ray Imaging Spectrometer (XIS; \citealt{Koyama2007}) 
was operated in the 
normal mode with no window option, providing a time resolution of 8~s. 
The Hard X-ray Detector (HXD; \citealt{Takahashi2007}) was in standard mode, 
with a time resolution of 61 $\mu$s for the individual events.

The XIS and HXD data were both processed with the \emph{Suzaku} 
pipeline version 2.4.12.27 in the \textsc{heasoft} (version 6.6) software package. 
The data were screened according to standard criteria. 
In particular, events were discarded if they were acquired during passages through 
the South Atlantic Anomaly or in regions of low geomagnetic cut-off rigidity 
($\le$6 GV for XIS and $\le$8 GV for HXD), or with low Earth elevation angles; 
for the XIS only events with grade 0, 2, 3, 4, and 6 were considered, 
and the \textsc{cleansis} script was used to remove hot or flickering pixels. 
The net exposures obtained with the XIS and HXD (live time 93.2\%) were 67.4 ks and 64.7 ks, respectively.

\begin{figure}
\begin{center}
\centerline{\includegraphics[width=9cm,angle=0]{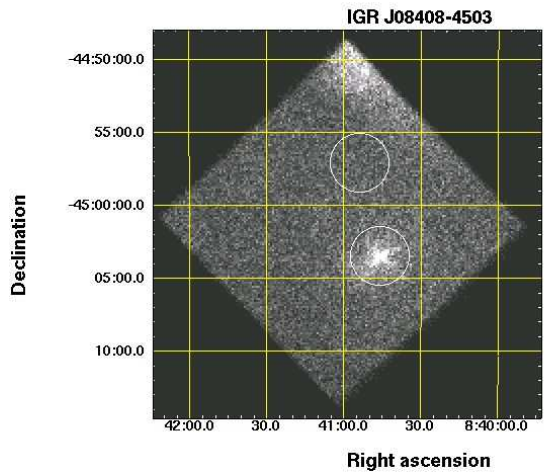}}
\caption{XIS1 image of the \src\ field. The two circular regions
mark the extraction regions for the source and the background.
In the upper part of the X--ray image, diffuse emission from the Vela SNR is evident (e.g., \citealt{lu2000}).
	}
\label{igr08408:fig:image}
\end{center}
\end{figure}

The XIS events of \src\ were accumulated in each of the three XIS cameras within a circular region (2$'$ radius) 
centered on the target, while the backgrounds were estimated from source-free regions 
well outside the point-spread function of the source (see Fig.~\ref{igr08408:fig:image}). 
For the spectral analysis, 
spectral redistribution matrices and ancillary response files for the 
XIS spectra were generated using the \textsc{xisrmfgen} and \textsc{xissimarfgen} tasks.

Background subtraction for the HXD data is performed using a synthetic model 
that accounts for the time-variable particle background (the so-called `non X-ray background', NXB). 
In particular, we used the `tuned' NXB model (see \citealt{Fukazawa2009} for details). 
The cosmic X--ray background and possible Galactic 
diffuse emission remain in the background subtracted data, 
as well as emission from any X-ray source that occurs in the 
instrument $34\times34$ arcmin$^2$ field of view.  

After subtracting the NXB, a positive signal is detected in the HXD-PIN up to $\sim$40 keV 
(about 10\% of the total counts in the 15--40 keV band); no significant emission is detected in the GSO data. 
To account for the expected contribution from the `cosmic X--ray background' (CXB), 
we used the spectra reported in \cite{Gruber1999} and in \cite{Moretti2009}. 
We estimate that the CXB contributes to the observed signal with 
roughly half of the NXB-subtracted counts.

The spectrum of the remaining signal is adequately described 
($\chi^2_\nu=1.16$ for 64 degrees of freedom)
 by a power-law model with photon index $1.9^{+0.8}_{-0.6}$ (1$\sigma$ confidence level) 
and 15--40 keV flux of $\approx$$6\times10^{-12}$ erg cm$^{-2}$ s$^{-1}$.
Considered that the combined uncertainties on the CXB and NXB are comparable 
to the flux estimated in this way, and lacking any specific signature of \src\ 
(such as it would be a periodic signal) in the HXD-PIN band, 
we regard the observed flux only as an upper limit on the hard X--ray emission of the source.
This upper limit is not very constraining for the extrapolation of the XIS spectrum
at high energy.
Thus, in the following, we concentrate on the soft energy part of the source spectrum observed with XIS.
%

\subsection{\inte\label{igr08408:interedu}}

The \emph{INTEGRAL} (INTErnational Gamma-Ray Astrophysics Laboratory, 
\citealt{Winkler2003}) 
satellite, launched in October 2002, carries 3 coded-mask telescopes:
\emph{IBIS} (15~keV- 10~MeV; \citealt{Ubertini2003}),
the spectrometer \emph{SPI} (20~keV$-$8~MeV; \citealt{Vedrenne2003}),
two X$-$ray monitors JEM-X1 and JEM-X2 (3$-$35~keV; \citealt{Lund2003}).
\emph{IBIS} is composed of two detector layers: ISGRI (15~keV - 1~MeV; \citealt{Lebrun2003}) 
and PICsIT (175~keV-10~MeV; \citealt{Labanti2003}).
An \emph{INTEGRAL} pointing lasts about 2~ks.

We analysed our \emph{IBIS/ISGRI} private data between 
55142.852 MJD and 55195.400 MJD, where \src\ was in the field of view 
(see Table \ref{integral_obs}),
using the Off-line Scientific Analysis package OSA~8.0 \citep{Goldwurm2003}.

\begin{figure}
\begin{center}
\centerline{\includegraphics[width=9cm,angle=0]{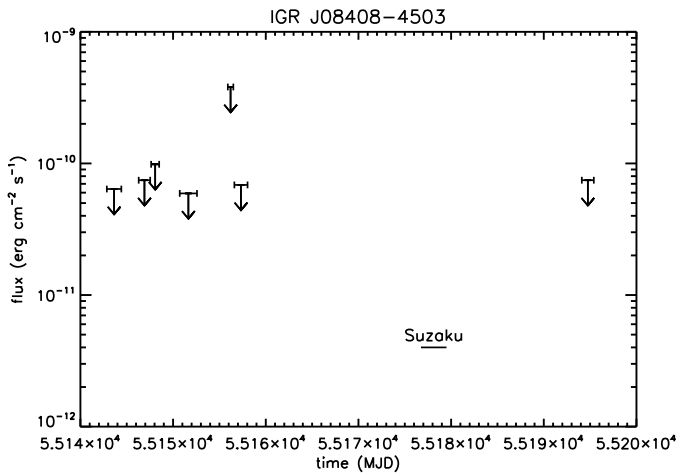}}
\caption{Long-term light curve of the 
\inte\ Key Program data of the source region, with the $5\sigma$ upper-limits obtained
with \emph{IBIS/ISGRI}, overimposed on the times of the \suz\ observation.
The \emph{IBIS/ISGRI} upper limits ($18-60$~keV) have been converted into flux in the 
 $1-100$~keV range, adopting the average
spectral parameters obtained by 
\citet{Romano2009:sfxts_paper08408}
	}
\label{igr08408:fig:intelc}
\end{center}
\end{figure}

\src\ has never been detected, being always below the $5\sigma$ threshold of detection, 
 both on timescales of a single observation, each lasting 2~ks, and of a single satellite 
revolution (Rev. in Table~1). 
The source remains undetected even adding together the whole data-set 
of the seven revolutions analysed here (Table~1).
In Figure \ref{igr08408:fig:intelc} we show the $5\sigma$ upper-limit lightcurve obtained
with \emph{IBIS/ISGRI} data, overimposed on the times of the \suz\ observation,
where we have converted the \emph{IBIS/ISGRI} count rate, measured in the 
$18-60$~keV range, to the $1-100$~keV unabsorbed flux, adopting the average
spectral parameters obtained by \citet{Romano2009:sfxts_paper08408}.

\begin{table}
\begin{center}
\caption{Summary of the \emph{INTEGRAL} data we have analysed.}
\label{integral_obs}
\begin{tabular}{lcc}
\hline
Rev. &       Period (MJD)       &                  Date                 \\
\hline
863  &   55142.852$-$55144.414  & 2009-11-07 20:26 $-$ 2009-11-09 09:56 \\
864  &   55146.289$-$55147.531  & 2009-11-11 06:56 $-$ 2009-11-12 12:44 \\
865  &   55147.637$-$55148.500  & 2009-11-12 15:17 $-$ 2009-11-13 12:00 \\
866  &   55150.715$-$55152.578  & 2009-11-15 17:09 $-$ 2009-11-17 13:52 \\
867  &   55155.910$-$55156.500  & 2009-11-20 21:50 $-$ 2009-11-21 12:00 \\
868  &   55156.613$-$55158.035  & 2009-11-21 14:42 $-$ 2009-11-23 00:50 \\
880  &   55194.098$-$55195.400  & 2009-12-29 02:21 $-$ 2009-12-30 09:36 \\
\hline
\end{tabular}
\end{center}
\end{table}

\subsection{\xmm\label{igr08408:xmmredu}}

We compare here our \suz\ results with a publicly available much shorter 
\xmm\ observation of a low luminosity state recently analysed by \citet{Bozzo2010}. 
We re-analysed this same data-set, to properly
compare it with our \suz\ observation
and check if the best fit found from \suz\ spectroscopy  equally well describes the \xmm\ spectrum,
at a similar source luminosity state.
\src\ was observed with \xmm\ on 2007 May 29, with a live time of 35.7~ks (EPIC pn). 

Data were reprocessed using version 9.0 of the Science Analysis
Software (SAS). Known hot, or flickering, pixels and electronic
noise were rejected. 
Response and ancillary matrix files
were generated using the SAS tasks {\em rmfgen} and {\em arfgen}.
We concentrate on the EPIC pn data: pn was operated in its Prime Full Window mode 
with all the CCDs in Imaging Mode.
The EPIC pn observation used the  thick filter.
Source pn spectrum was extracted from a circular region
of 30\arcsec\ radius to avoid the edge of the CCD. 
We selected PATTERNS from 0 to 4.
Background counts were obtained from similar sized region offset from the source position. 
The background (selected with PATTERN=0 and above 10 keV) 
showed evidence of some flaring activity, which was filtered out, resulting in a net exposure
time for the final source pn spectrum of 35.1~ks.
The \src\ pn spectrum 
was not affected by pile-up, 
resulting in an average net count rate of 0.170$\pm{0.003}$~s$^{-1}$.

To ensure applicability of the \chisq\ statistics, the
net spectra were rebinned such that at least 30 counts per
bin were present.
All quoted uncertainties are given at 90\% confidence level for 
one interesting parameter.

  	\section{Analysis and Results\label{igr08408:res}}

  	\subsection{\suz\label{igr08408:suzaku}}

\begin{figure*}
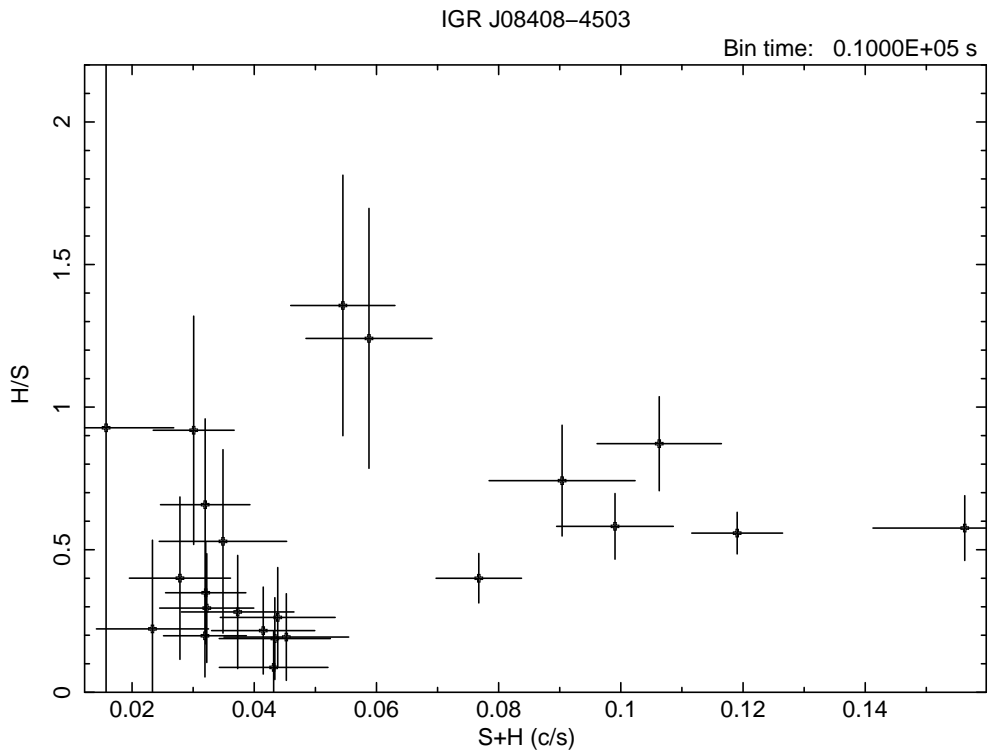

\begin{center}
\begin{tabular}{cccc}
\includegraphics[height=13.5cm,angle=-90]{lsidoli_fig04a.ps} \\
\includegraphics[height=13.3cm,angle=-90]{lsidoli_fig04b.ps}
\end{tabular}
\end{center}
\caption{\suz/XIS1 net light curves of \src\ in two energy ranges 
(the XIS1 count rate S is the background subtracted source rate in the range 0.4-3 keV, while
H is the net count rate in the range 3-10 keV), together with the hardness ratio (H/S) versus time 
and versus the intensity (S+H). Bin time is 5~ks in the upper panels, while 10~ks in the lower panel.
Start and stop times are in units of MJD - 40,000. }
\label{igr08408:fig:x1lc}
\end{figure*}

The background subtracted light curve of \src\ in two energy bands (softer and harder than 3 keV) together
with their hardness ratio, are shown in Fig.~\ref{igr08408:fig:x1lc}.
The hardness ratio does not show strong
evidence of variability, thus we will study both the average spectrum 
extracted from the
whole exposure time, as well as time selected spectra from the initial part of the
observation and from the peak of the two flares, separately.
Fitting the whole hardness ratio as a function of time (with a bin time of 5~ks) 
to a constant model
we obtain a value of $0.40\pm0.03$ ($\chi^2_{\nu}=1.3$ for 44 d.o.f.).  

A timing analysis was performed   
after having converted the event arrival times to the Solar System 
Barycentric frame. We searched for coherent periodicity, 
but found no evidence for pulsations. 
We could place an upper limit on the pulsed fraction 
(defined as the semi-amplitude of the sinusoidal modulation devided the mean count rate), computed according 
to \citet{Vaughan1994}, of 40\% at the 99\% confidence level 
for periods in the range 16\,s--500\,s.

We selected three spectra from  the \suz\ data-set, as follows:
the first spectrum (which we call ``persistent'' spectrum in Table~\ref{igr08408:tab:spec})
was extracted accumulating all photons coming 
from the first 120~ks of the observation, with the lowest intensity level
(see the light curve reported in Fig.~\ref{igr08408:fig:x1lc}); 
the second spectrum is from the peak of the 
first flare 
(``flare 1'', between 145~ks and 160~ks 
in Fig.~\ref{igr08408:fig:x1lc}),
while the third spectrum comes from the peak of the 
second flare (``flare 2'',  between 202~ks and 216~ks 
in Fig.~\ref{igr08408:fig:x1lc}).
These time spans are subjective choices
in order to extract the lowest intensity spectrum and the spectra at the peak of each flare. 

We first analysed the lowest intensity  spectrum (the ``persistent'' pre-flare spectrum, 
with a net exposure time of 35~ks), 
always fitting together the spectra
from the different XIS units (extracted separately), with factors included in 
the spectral fitting to allow for normalization 
uncertainties between the instruments. 
A fit with an  absorbed simple power law ({\sc pegpwrlw} in {\sc xspec})
resulted in positive residuals below 1~keV and in an unphysical very low energy 
absorption consistent with zero (resulting in a photon index of 1.5, with a reduced 
$\chi^{2}_{\nu}$/d.o.f.=1.248/57).
Thus in the following we will constrain the absorbing column density to be at least  
that predicted from optical measurements 
towards the donor star ($N_{\rm H}$=3$\times$$10^{21}$ cm$^{-2}$; \citealt{Leyder2007}).
We then added a second component to the power law continuum
(a thermal plasma model,  {\sc mekal} model in {\sc xspec}), 
to account for the residuals evident at soft energies. 
We note that a similar  low energy component was also found in the low intensity state observed in 2007
with \xmm\ \citep{Bozzo2010} and explained by these authors as the likely 
X--ray emission from the blue supergiant companion.
The double component model resulted in a much better fit ($\chi^{2}_{\nu}$/d.o.f.=0.811/55), with
the spectral parameters reported in Table~\ref{igr08408:tab:spec}.
If we extrapolate at high energies this best fit, we derive a flux of 
4.35$\times$$10^{-13}$ erg cm$^{-2}$ s$^{-1}$ (15--40 keV),
more than one order of magnitude lower than the upper limit 
we can place with \suz/HXD (see Sect.~\ref{igr08408:suzredu}).
If interpreted as  X--ray emission from the companion star, this soft component
should be obviously always present with similar spectral parameters, 
and with an X--ray luminosity
consistent with that usually observed from OB supergiants 
($\sim$10$^{31}$--10$^{32}$~erg~$^{-1}$; \citealt{Cassinelli1981}).
Thus we included it in the fitting model 
also when analysing the other two spectra from the peak of the two flares.

Fitting the first flare peak emission with this double component model 
(net exposure time of 7.8~ks; Table~\ref{igr08408:tab:spec}),
we found a good fit ($\chi^{2}_{\nu}$/d.o.f.= $1.041/71$) 
but with a very low temperature $kT_{\rm mekal}$ $<$0.09 keV 
and a high normalization 
for this soft component, which
translates into an X--ray luminosity (0.5--10~keV) of 4$\times$10$^{35}$~erg~$^{-1}$,
too high to be compatible with the X--ray emission from the supergiant companion.
Thus, we constrained the {\sc mekal} temperature and its normalization to be within
the same variability ranges we obtained from the spectroscopy 
of the faint persistent spectrum  (Table~\ref{igr08408:tab:spec}), which we verified to be also consistent with
the faintest emission observed in 2007 with \xmm, and reanalysed here in Sect.~\ref{igr08408:xmm}.
Constraining these two parameters in the double component continuum, we obtained 
a worse fit and wave-like residuals ($\chi^{2}_{\nu}$/d.o.f.= $1.355/71$) along the whole energy range.
To solve this problem 
we added an  additional absorption to the power law model, which can be explained 
with a different absorption towards the donor star and
the compact object.
This new double component model 
({\sc phabs} $\times$ ({\sc mekal} +{\sc phabs} $\times$ {\sc pegpwrlw}) adopting the {\sc xspec} syntax)
resulted in the best fit ($\chi^{2}_{\nu}$/d.o.f.= $0.958/70$) 
reported in Table~\ref{igr08408:tab:spec},
with an additional absorption, $N_{\rm H pow}$, towards the power law component of $\sim$1.8$\times$$10^{22}$ cm$^{-2}$
and a power-law photon index, $\Gamma$, of $\sim$2.1.
The unabsorbed flux reported in Table~\ref{igr08408:tab:spec} is the flux obtained correcting from the
total absorption $N_{\rm H}$, but not for $N_{\rm H pow}$. 
Correcting also for $N_{\rm H pow}$, and measuring the
unabsorbed flux coming only from the power law component, we obtained 
6$\times$$10^{-12}$~erg~cm$^{-2}$~s$^{-1}$ (1--10 keV).

Fitting the second flare peak emission with this same double component model 
(net exposure time of 2.7~ks; see Table~\ref{igr08408:tab:spec} for the fit results),
we obtained   $N_{\rm H pow}$$\sim$1.4$\times$$10^{22}$ cm$^{-2}$ for the power law component, 
consistent
with that measured during the first flare.
The unabsorbed flux measured from only the power law
component is 7$\times$$10^{-12}$~erg~cm$^{-2}$~s$^{-1}$ (1--10 keV) which, extrapolated to the
range 15--40 keV, results into an unabsorbed flux of 3$\times$$10^{-12}$~erg~cm$^{-2}$~s$^{-1}$,
consistent with the upper limit placed from HXD observation.

 \begin{table*}
 \begin{center}
 \caption{Results of the time resolved spectroscopy (persistent, flare 1 and flare 2) and of the 
average spectrum extracted from the whole \suz/XIS observation. \label{igr08408:tab:spec} }
 \begin{tabular}{lllcrrcc}
 \hline
 \hline
 \noalign{\smallskip}
 Spectrum & $N_{\rm H}$ 		& $N_{\rm H pow}$        & $\Gamma$ 	&  $kT_{\rm mekal}$ & $norm_{\rm mekal}$ & Unabs Flux  (1-10 keV)    & $\chi^{2}_{\nu}$/d.o.f.\\
          & ($10^{22}$ cm$^{-2}$)& ($10^{22}$ cm$^{-2}$)  &	        &  (keV) 	   &    ($10^{-4}$)   &   ($10^{-12}$ erg cm$^{-2}$ s$^{-1}$) &           \\
 \noalign{\smallskip}
 \hline
 \noalign{\smallskip}
Persistent        &   $0.30_{-0.0}^{+0.20}$  & $-$                   & $1.57_{-0.17}^{+0.18}$  & $0.22_{-0.04}^{+0.07}$   & $4.7_{-2.7}^{+18.9}$ &  0.42  & $0.811/55$  \\
                  &   $0.30_{-0.0}^{+0.24}$ & $0.63_{-0.52}^{+0.63}$   & $2.02_{-0.41}^{+0.50}$  &  $0.23 _{-0.06}^{+0.07}$  &  $4.9  _{-2.0}^{+24.0}$ & 0.44 & $0.752/54$  \\
Flare 1           &   $2.19_{-0.45}^{+0.45}$ & $-$                   & $1.95_{-0.21}^{+0.21}$  &  $<$0.09                 &  $100_{-70}^{+160}$  & 9.2    & $1.041/71$  \\
                  &   $0.77_{-0.47}^{+0.25}$ & $1.78_{-0.42}^{+0.53}$  & $2.08_{-0.26}^{+0.25}$  &  $0.25_{-0.07}^{+0.04}$    &  $23 _{-21}^{+0}$    & 4.0    & $0.958/70$  \\
Flare 2           &   $0.85_{-0.35}^{+0.18}$ & $-$                   & $1.40_{-0.18}^{+0.18}$  &  $0.28 _{-0.07}^{+0.01}$   &  $23  _{-21}^{+0}$   & 5.8   & $1.869/31$   \\
                  &   $0.61_{-0.18}^{+0.12}$ & $1.38_{-0.75}^{+0.64}$  & $2.00_{-0.33}^{+0.36}$  &  $0.29 _{-0.11}^{+0.0}$    &  $23  _{-21}^{+0}$   & 5.2   & $1.492/30$  \\
 \noalign{\smallskip}
 \hline
 \noalign{\smallskip} 
Average           &   $0.62_{-0.06}^{+0.20}$ & $-$                   & $1.39_{-0.06}^{+0.07}$  &  $0.18 _{-0.0}^{+0.02}$   &  $33  _{-18}^{+15}$   & 1.5   & $1.646/265$   \\
                  &   $0.47_{-0.17}^{+0.13}$ & $1.28_{-0.23}^{+0.24}$ & $1.98_{-0.13}^{+0.14}$  &  $0.23 _{-0.04}^{+0.07}$    &  $11  _{-8}^{+18}$  & 1.4   & $1.211/264$  \\
  \noalign{\smallskip}
  \hline
  \end{tabular}
  \end{center}
  \end{table*}

\begin{figure}
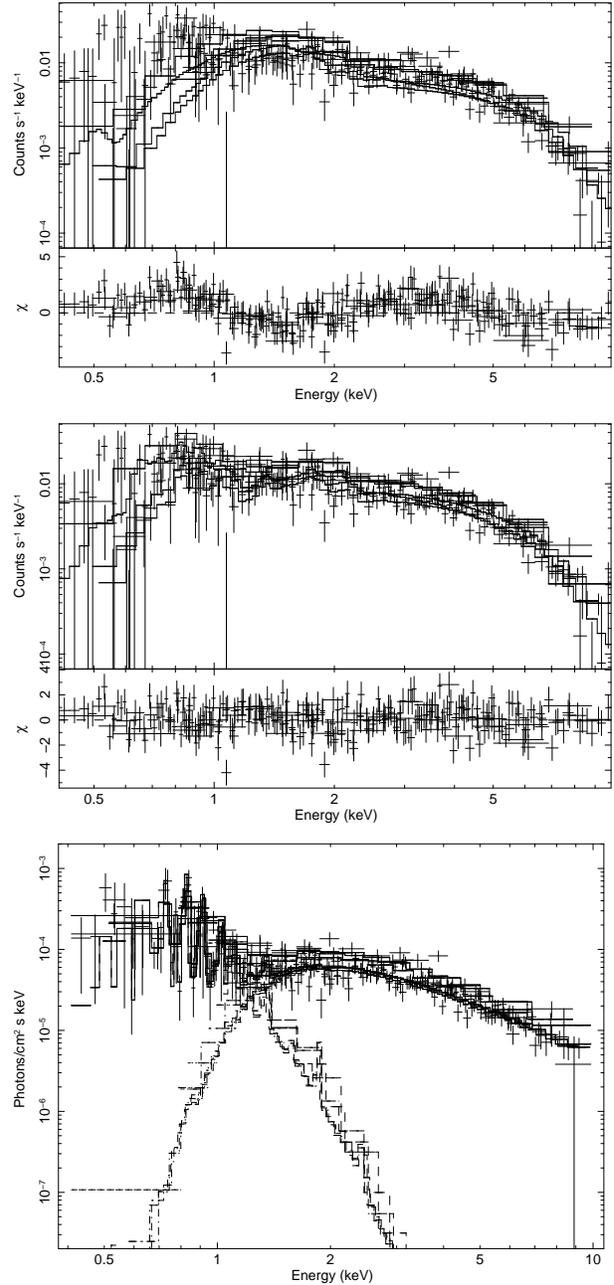

\begin{center}
\begin{tabular}{cccc}
\includegraphics[height=8.0cm,angle=-90]{lsidoli_fig05a.ps} \\
\includegraphics[height=8.0cm,angle=-90]{lsidoli_fig05b.ps} \\
\includegraphics[height=8.0cm,angle=-90]{lsidoli_fig05c.ps}
\end{tabular}
\end{center}
\caption{Spectral fits to the \suz\ spectrum of \src\ extracted from the whole observation.
The {\em upper panel} shows the result adopting an absorbed power law model (counts spectrum
together with the wave-like residuals in units of standard deviation).
The {\em medium and lower panels} display the best fit to the average \src\ spectrum, obtained with
a double component model composed by a {\sc mekal} 
at low energies together with a differently absorbed powerlaw model ({\sc pegpwrlw} in
{\sc xspec}).
}
\label{igr08408:fig:avspec}
\end{figure}

Finally, we performed a spectral fit to the average spectrum extracted from the whole \suz\ observation
(net exposure time 67.4~ks). 
The results obtained with the same double component model discussed before
are reported in Table~\ref{igr08408:tab:spec} and shown in Fig.~\ref{igr08408:fig:avspec}.

  	\subsection{\xmm \label{igr08408:xmm}}

The source low intensity state in the \suz\ observation is similar to that observed
with \xmm\ in 2007. 
Thus, we re-analysed here these same public EPIC pn data, in order to compare them
with our \suz\ XIS results. 
In particular, we  performed a different
time selection spectroscopy with respect to that reported by Bozzo et al. (2010).
We then reanalysed the spectra adopting the same best-fit model found from \suz\ analysis,
to test if the same double component spectral model can explain both data-sets.

\begin{figure}
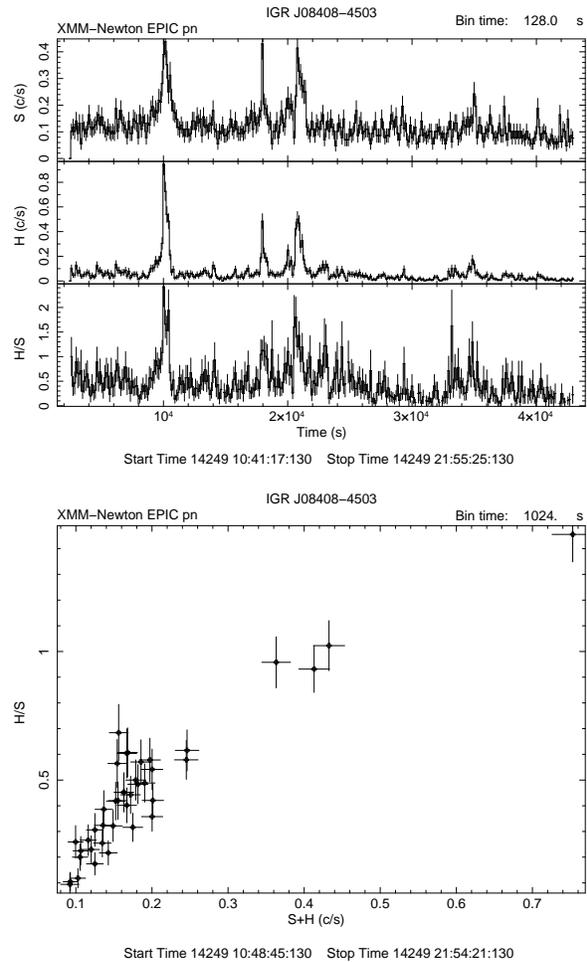

\begin{center}
\begin{tabular}{cccc}
\includegraphics[height=8.0cm,angle=-90]{lsidoli_fig06a.ps} \\
\includegraphics[height=8.0cm,angle=-90]{lsidoli_fig06b.ps}
\end{tabular}
\end{center}
\caption{\xmm\ EPIC pn light curves of \src\ in 2007, in two energy ranges 
(S is the total source rate in the range 0.2-2 keV, while
H is the total count rate in the range 2-12 keV), together with the hardness ratio (H/S) versus time 
and versus the intensity (S+H). Bin time is 128~s in the upper panels, while 1024~s in the lower panel.
Start and stop times are in units of MJD - 40,000. }
\label{igr08408:fig:xmm}
\end{figure}

The light curve observed in 2007 with \xmm\ (EPIC pn data) is reported in
Fig.~\ref{igr08408:fig:xmm}, in two energy ranges (above and below 2 keV) together with
their hardness ratio. In the same figure (lower panel) we also show
an hardness-intensity plot, with a larger bin time, to clearly show
that there is a clear correlation between the hardness ratio and the source intensity,
as already noted and discussed by \cite{Bozzo2010}.

Since hardness ratio and source intensity are correlated, an intensity-selected spectroscopy
translates also into a hardness-ratio selected spectroscopy.
In order to compare the \suz\ spectrum of the initial lowest intensity state (``persistent'' emission
in Table~\ref{igr08408:tab:spec}) with
the lowest intensity state observed with \xmm, we extracted a spectrum where the
EPIC pn rate (S+H in Fig.\ref{igr08408:fig:xmm}, lower panel) 
was lower than 0.14~s$^{-1}$ (spectrum A in Table~\ref{igr08408:tab:specxmm}). 
During this low intensity, the contribution to the X--ray emission
from the O-type supergiant companion is maximum, thus allowing us to put 
better constraints to the companion soft energy component (e.g. \citealt{Cassinelli1981}).
 Three other  intensity-selected spectra were extracted in the following EPIC pn total rate ranges
(S+H in Fig.\ref{igr08408:fig:xmm}, lower panel): 
0.14~s$^{-1}$--0.3~s$^{-1}$ (spectrum B in Table~\ref{igr08408:tab:specxmm}), 
0.3~s$^{-1}$--0.7~s$^{-1}$ (spectrum C)
and $>$0.7~s$^{-1}$, to catch only the peaks of the flaring activity (spectrum D).

We applied to the pn intensity-selected spectra the best fit model found from
\suz\ data, a double component model with a continuum composed by a hot plasma model
together with a power law model, differently absorbed 
({\sc phabs} $\times$ ({\sc mekal} +{\sc phabs} $\times$ {\sc pegpwrlw}) adopting the {\sc xspec} syntax).
We constrained the total absorption (the first {\sc phabs} model, absorbing both continuum components)
to be not less than the absorption towards the optical companion (as during the \suz\ XIS analysis).
We also constrained the {\sc mekal} parameters to be within the ranges found from the \suz\ XIS analysis.
In Table~\ref{igr08408:tab:specxmm} the spectral results of only the power law component with the
total and the additional absorption are reported.
The fit results are good, demonstrating that the same model adopted to explain
\suz\ spectra is a nice deconvolution of the \xmm\ spectra too, 
although  a simpler model (an absorbed power law) 
already results in a good
description of the spectra A and D.
We note that the main responsible for the increasing of the hardness ratio during flares
is very likely a harder power law high energy component rather than an increasing of the
absorbing column density (which remains constant, within the uncertainties), as already reported
by Bozzo et al. (2010), but adopting a different deconvolution of the spectrum.

 \begin{table*}
 \begin{center}
 \caption{\xmm\ results of the source intensity (and hardness) selected spectroscopy (EPIC pn) adopting
the model {\sc phabs} $\times$ ({\sc mekal} +{\sc phabs} $\times$ {\sc pegpwrlw}). 
See Sect.~\ref{igr08408:xmm} for details. 
 \label{igr08408:tab:specxmm} }
 \begin{tabular}{lllccc}
 \hline
 \hline
 \noalign{\smallskip}
 Spectrum & $N_{\rm H}$ 		& $N_{\rm H pow}$        & $\Gamma$ 	 & Observed Flux  (1-10 keV)    & $\chi^{2}_{\nu}$/d.o.f.\\
          & ($10^{22}$ cm$^{-2}$)& ($10^{22}$ cm$^{-2}$)  &	         &   ($10^{-13}$ erg cm$^{-2}$ s$^{-1}$) &           \\
 \noalign{\smallskip}
 \hline
 \noalign{\smallskip}
A     &   $0.67_{-0.37}^{+0.06}$ & $<$2.1                &  $2.94_{-0.69}^{+0.92}$      & 1.8  & $1.209/20$  \\
B     &   $0.61_{-0.15}^{+0.13}$ & $0.89_{-0.25}^{+0.29}$  &  $2.42_{-0.19}^{+0.21}$     & 6.4  & $1.090/64$  \\        
C     &   $0.49_{-0.18}^{+0.33}$ & $0.81_{-0.32}^{+0.38}$  &  $1.95 _{-0.23}^{+0.25}$    & 23   & $0.993/19$  \\
D     &   $0.50_{-0.10}^{+0.10}$ & $0.51_{-0.45}^{+0.50}$  &  $1.46 _{-0.27}^{+0.25}$    & 58   & $0.713/11$  \\
  \noalign{\smallskip}
  \hline
  \end{tabular}
  \end{center}
  \end{table*}

	\section{Discussion and Conclusions\label{igr08408:discussion}}

We report here on the longest observation of a low luminosity state
of a SFXT, spanning about three consecutive days (with the obvious gaps due to
the satellite orbits), resulting in a net exposure time of 67.4~ks.
The \suz\ \src\ observation unveils a source luminosity level at about  
4$\times$10$^{32}$~erg~s$^{-1}$ for the first $\sim$120~ks, 
followed by two long flares (with a duration of about 45~ks each) with a count rate a factor of $\sim$3 higher 
than the  initial ``persistent'' pre-flare emission.
The source has not been convincingly detected with HXD and we could place 
the best upper limit to date to the hard X--ray emission (15--40 keV)
of a SFXT in the low intensity state, at a level of 6$\times$10$^{33}$~erg~s$^{-1}$.

The time selected spectroscopy of the XIS observation resulted 
in a best fit model composed by a double component continuum (a soft hot plasma
model together with a power law dominating at high energies), differently absorbed, with 
the presence of an extra absorption of the high energy power law component. 
This additional absorbing matter, of the order of $\sim$1--2$\times$10$^{22}$ cm$^{-2}$,
is very likely local to the compact object, since it absorbs only the power law component. 
It could be due to dense wind clumps passing in front of the compact object and close to it, 
or more likely  to the accreting matter itself, since this additional absorption
is more clearly detected during the two long duration flares. 
The spectral parameters of the double component model suggest 
that the low energy soft component
is produced by the supergiant companion, while the high energy component
is too hard for a stellar origin (we will not repeat here the convincing 
arguments already discussed by Bozzo et al. 2010).
Both the temporal and spectral properties indicate that in \src\ the compact object
is still accreting matter even at the low intensity state we observed with \suz: indeed,
also in the lowest luminosity state of 4$\times$10$^{32}$~erg~s$^{-1}$,
during the first part of the observation (long term pre-flare emission), 
the \suz\ spectrum is hard and power-law-like at energies 
higher than 1 keV, with a photon index, $\Gamma$, of 1--2.
Similar photon indexes have been measured in the 1--10 keV spectra of  four other  SFXTs 
monitored with \sw\ during their long-term intermediate states \citep{Sidoli2008:sfxts_paperI},
although at higher luminosities (10$^{33}$--10$^{34}$~erg~s$^{-1}$).
The presence of flares, together with the hard X--ray emission, is strongly indicative
of the fact that accretion on the compact object  is still ongoing in \src, even at very low luminosities.

SFXTs behaviour is still not explained (both the outbursts and the high dynamic range), 
although several possible 
mechanisms have 
been proposed since the times of their discovery with \inte\ \citep{Sguera2005}.
Among the different proposed explanations (see \citealt{Sidoli2009:cospar} for a review) 
which involve the wind properties and/or the characteristics of the compact object
 (neutron star magnetic field and spin period) 
there is a general consensus on the fact 
that the supergiant wind is very likely not homogeneous (``clumpy''; \citealt{zand2005}).

In the framework of the clumpy wind model developed by \citet{Ducci2009},
we can derive the distance of the neutron star from the supergiant companion,
from the durations and luminosities of the flares observed with \suz,
and adopting the expansion law for the clump size 
(see Eq. [9] in \citealt{Ducci2009}; see also \citealt{Romano2009:sfxts_paper08408} 
and \citealt{Sidoli2009}
for previous applications of this method to \src).
Assuming a wind velocity of $v_{\rm w}=1800$~km~s$^{-1}$
and a neutron star with mass $M_{\rm x}=1.4$~M$_\odot$,
the accretion radius is:
\begin{equation} \label{racc}
R_{\rm a} = \frac{2 G M_\odot}{v_{\rm w}^2} \approx 1.2 \times 10^{10} \mbox{ cm}  
\end{equation}
From the time duration of the flares, $t_{\rm fl} \simeq 4.5 \times 10^{4}$~s,
we obtain the radius of the clumps accreted by the neutron star:
\begin{equation} \label{rcl}
R_{\rm cl} = \frac{v_{\rm w} t_{\rm fl}}{2} \approx 4 \times 10^{12} \mbox{ cm} 
\end{equation}

Extrapolating flux of the two flares to a broader energy range, 
$F_{\rm x} \simeq 1 \times 10^{-11}$~erg~cm$^{-2}$~s$^{-1}$ ($1-40$~keV),
and assuming a distance of $d=3$~kpc,
we obtain a flare X$-$ray luminosity $L_{\rm x} \simeq 10^{34}$~erg~s$^{-1}$.
Hence, from the accretion X--ray luminosity
and the continuity equation $\dot{M}_{\rm accr} = \rho v_{\rm w} \pi R_{\rm a}^2$,
we obtain a clump density of $\rho_{\rm cl} \simeq 6.6 \times 10^{-16}$~g~cm$^{-3}$,
and a mass of $M_{\rm cl} \approx 10^{23}$~g.
For the supergiant companion we assumed the following physical properties: 
a radius $R_{\rm OB}=23.8$~R$_\odot$, 
a mass $M_{\rm OB}=30$~M$_\odot$, a luminosity  $\log L/L_\odot = 5.847$ 
and an effective temperature $T_{\rm eff} = 34000$~K \citep{Vacca1996}. 
We thus obtain, from the clump mass-radius equation of Ducci et al. 2009 (Eq. [20]),
the starting radius of the clump: $R_{\rm cl,i} \simeq 1.7 \times 10^{11}$~cm.
Assuming that the sonic radius $R_{\rm s}$ equals the supergiant radius, 
and a starting wind velocity at $R_{\rm s}$ 
of $v_{\rm w}(R_{\rm s}) \simeq 3 \times 10^6$~cm~s$^{-1}$ \citep{Bouret2005}, 
we obtain, from the expansion law of the clump, 
a distance of the neutron star of $r \simeq 10^{13}$~cm,
consistent with our previous findings \citep{Romano2009:sfxts_paper08408}.

The orbital separation derived here implies an orbital period of about 35~days, which is 
somehow intermediate between the narrow orbits of other SFXTs
(e.g., 3.3~days in IGR~J16479-4514, \citealt{Jain2009:16479}) and the longest orbital
period measeured to date in a SFXT (165~days, \citealt{SidoliPM2006}, \citealt{Romano2009:11215_2008}), whereas 
it is very similar to that measured in SFXTs like, e.g,  
SAX~J1818.6--1703 (\citealt{Zurita2009}, \citealt{Bird2009}).  
Even the long duration low luminosity emission at a few 10$^{32}$~erg~s$^{-1}$ (close to the
quiescent emission observed, e.g., in IGR~J17544--2619, \citealt{zand2005}) is consistent
with a wide orbit, in order to allow accretion at a low level. 
Also the 
long term light curve reported here from \inte\ observations
seems to point to a long duration of the low intensity emission in \src.

The presence of short flaring activity as observed with \xmm\ 
in \src\ during a low intensity state 
has already been caught in other SFXTs (e.g., in IGR J17544-2619, \citealt{Gonzalez2004}, their Fig.~3)
with flares durations of a few hundred seconds.
On the other hand, the \suz\ pointing spanning three days allowed us for the
first time  to probe the source light curve on much longer timescales,
catching longer flares ($\sim$45~ks) than usual, with
peak rates which are only $\sim$3 times higher than the average pre-flare intensity.
The observed properties of these flares, their long duration 
together with the low dynamic range with respect to the
pre-flare emission, suggested us to search for other possible explanations, although in the
framework of the structure of the supergiant wind.
A viable alternative to the accretion of single clumps 
could be that the flares are produced from 
the accretion of large scale wind structures, like gas streams, or  
the so-called corotating interaction regions (CIRs, \citealt{Mullan1984}).
CIRs form in a stellar wind
when the rotating star emits wind in a non-spherically symmetric manner. 
These wind structures were originally studied in the solar corona  \citep{Mullan1984}
and form as interaction regions between slow and fast streamlines. 
They were later 
suggested to be present also in rotating O-star winds \citep{Cranmer1996}, 
with a mild density contrast of a factor of
2--3 with respect to the undisturbed wind.
These structures have a typical thickness of $\sim$0.1$r$ at radial distance $r$ \citep{Mullan1984}.
Thus, assuming a radial distance $r \simeq 10^{13}$~cm and a stream thickness, $\Delta$S$\sim$0.1$r$,
we can derive the neutron star velocity, v$_{ns}$ from $\Delta$S and the flare duration $\Delta$t$_{flare}$
($\Delta$S=v$_{ns}$$\times$ $\Delta$t$_{flare}$). 
Assuming $\Delta$t$_{flare}$=45~ks and $\Delta$S$\sim$0.1$r$=$10^{12}$~cm,
we find  v$_{ns}$=200~km~s$^{-1}$, which is consistent with 
the neutron star orbital velocity in a 35~days orbit
around a 30~M$_\odot$ companion star.

\section*{Acknowledgments}

This work was supported in Italy by contract ASI/INAF I/088/06/0.
P.E. acknowledges financial support from the Autonomous Region of Sardinia 
through a research grant under the program PO Sardegna FSE 2007--2013, 
L.R. 7/2007 ``Promoting scientific research and innovation technology in 
Sardinia''.
This work is based on data from observations with \suz, \inte, and \xmm.
\suz\ is a Japan's mission developed at the Institute 
of Space and Astronautical Science of Japan Aerospace Exploration Agency 
in collaboration with U.S. (NASA/GSFC, MIT) and Japanese institutions.
\xmm\ is an ESA science mission with instruments and 
contributions directly funded by ESA Member States and the USA (NASA). 
\inte\ is an ESA project with instruments and 
science data centre funded by ESA member states 
(especially the PI countries: Denmark, France, Germany, Italy, Switzerland, Spain), 
Czech Republic and Poland, and with the participation 
of Russia and the USA. 
This research has made use of HEASARC online services, supported by NASA/GSFC.


\bsp

\label{lastpage}

\end{document}